\documentclass[%
reprint,
 amsmath,amssymb,
 aps, physrev
]{revtex4-2}

\usepackage{graphicx}
\usepackage{dcolumn}
\usepackage{bm}
\usepackage{hyperref}
\usepackage{xcolor}

\begin{document}

\preprint{APS/123-QED}

\title{\textbf{Memory-induced current reversal of Brownian motors} 
}%

\newcommand{\todo}[1]{\textbf{\color{red}#1}}

\author{Mateusz Wi\'{s}niewski}
\author{Jakub Spiechowicz}%
 \email{jakub.spiechowicz@us.edu.pl}
\affiliation{%
    Institute of Physics, University of Silesia, 41-500 Chorz\'{o}w, Poland
}%


\begin{abstract}
Kinetics of biological motors such as kinesin or dynein is notably influenced by viscoelastic intracellular environment. The characteristic relaxation time of the cytosol is not separable from the colloidal timescale and therefore their dynamics is inherently non-Markovian. In this paper we consider a variant of a Brownian motor model, namely a Brownian ratchet immersed in a correlated thermal bath and analyze how memory influences its dynamics. In particular, we demonstrate the memory-induced current reversal effect and explain this phenomenon by applying the effective mass approximation as well as uncovering the memory-induced dynamical localization of the motor trajectories in the phase space. Our results reveal new aspects of the role of memory in microscopic systems out of thermal equilibrium.

\end{abstract}

\maketitle


\section{\label{sec:intro}Introduction}

Dynamics of small systems in microscopic world is noticeably affected by thermal fluctuations, the influence of which in macroscopic reality is typically rather a nuance. It is tempting to use this inevitable noise as an energy source and make useful work out of it. Although the second law of thermodynamics prohibits rectifying thermal fluctuations in equilibrium \cite{Smoluchowski1912, Feynmann1970}, this limitation is abandoned in nonequilibrium. In fact this is the \emph{modus operandi} of many natural microscopic machines such as e.g. molecular motors \cite{Peskin1993, Julicher1997, Bressloff2013, Kay2015, Astumian2016, Hoffmann2016, Borsley2024, Burnham2019, McCausland2021, Mazal2021}.

A paradigmatic model of a molecular motor is a Brownian motor, a system with broken spatio-temporal symmetry consisting of a Brownian particle dwelling in a periodic potential \cite{Astumian1997, Astumian2002, Reimann2002a, Hanggi2009, Denisov2014, Cubero2016}. The first minimal models of Brownian motors emerged in physics in early 1990s \cite{Ajdari1992, Magnasco1993, Ajdari1994, Bartussek1994, Astumian1994, Doering1994, Millonas1994} and were soon followed by pioneering experiments \cite{Rousselet1994, Faucheux1995} corroborating the theoretical predictions. 

Despite more than three decades of research the Brownian motor concept has not lost its prominence but instead it is even expanding and nowadays is discovered in systems like colloidal particles \cite{vanOudenaarden1999, Matthias2003, Tierno2014, Arzola2017, Schwemmer2018, Skaug2018, Leyva2022, Camacho2023}, superconductors \cite{Wambaugh1999, Olson2001, Savelev2002, Sterck2002, Villegas2003, Hastings2003, Zhu2004, deSouzaSilva2006}, molecular walkers \cite{vonDelius2009, Barrell2010}, molecular rotary motors \cite{Leigh2003, Pumm2022}, DNA transport machinery \cite{Molcrette2022}, cold atoms in optical lattices \cite{Gommers2005, Gommers2006, Dupont2023}, magnetic domain walls \cite{Perez-Junquera2008, Franken2012}, dusty plasma \cite{He2020}, active matter \cite{Angelani2009, DiLeonardo2010, Reichhardt2017, Granek2022, Patil2022}, crystalline materials \cite{Qiu2024} and granular gas \cite{Lagoin2022}, to name only a few.

A symmetry-breaking mechanism may be intrinsically embedded in a Brownian motor in absence of any perturbations. This case known as a Brownian ratchet is typically realized as an asymmetric spatially periodic potential. There exist a variety of ratchet types depending on the origin of nonequilibrium state necessary to overcome limitations of the second law of thermodynamics including e.g. rocking \cite{Magnasco1993, Bartussek1994, Jung1996, Zapata1996, Mateos2000, Machura2004, Spiechowicz2014, Spiechowicz2015a, Spiechowicz2017a}, pulsating \cite{Astumian1994, Chauwin1995, Bier1996a} and correlation ones \cite{Doering1994, Millonas1994, Luczka1995, Bartussek1996, Kula1998a}, for more details see \cite{Hanggi2009, Cubero2016}. 

A consequence of the Curie's principle \cite{Curie1894}, telling that if a certain phenomenon is not ruled out by symmetry then it would emerge, is that an average velocity of a Brownian motor may be non-zero even in spite of the absence of any biased forces or gradients. However, the direction of transport is typically difficult to predict \emph{a priori} and an effect of current reversal can be observed upon variation of system parameters characterizing potential \cite{Chauwin1995, Bier1996a, Jiao2023}, external force \cite{Jung1996, Mateos2000, Machura2004, Cubero2010}, thermal noise intensity \cite{Bartussek1994, Kula1998a, Machura2004}, nonequilibrium fluctuations \cite{Millonas1994, Doering1994}, particle mass \cite{Marchesoni1998, Lindner1999} or friction \cite{Bier1996a, Gnoli2013}. This fact, in turn, opens the possibility for transport control and its various applications.

Biological motors, e.g. kinesin or dynein, that are precursors of artificial Brownian motors, operate in viscoelastic intracellular environment in which the characteristic relaxation time scales for the fluid and motor are comparable. In such a case memory effects are no longer negligible and the system dynamics is non-Markovian. Although the influence of memory on the kinetics of Brownian motors has been already addressed \cite{Goychuk2010, Kharchenko2013, Goychuk2014a} its impact on the emergence of current reversal phenomenon has not been investigated yet. In this work we fill this gap and report the memory-induced current reversal effect. In doing so we analyze a rocking ratchet system immersed in a correlated thermal bath and show that this phenomenon emerges upon increasing the memory time of the setup while it ceases to exist in the memoryless situation. 

The paper is organized as follows. In Sec.~\ref{sec:model} we detail on the studied model together with the employed methods. Then in Sec.~\ref{sec:results} we demonstrate the memory-induced current reversal effect and explain its mechanism by applying the effective mass approach \cite{Wisniewski2024-effmass, Wisniewski2024-anm}. Finally, Sec.~\ref{sec:conclusions}. provides a summary and conclusions.

\section{\label{sec:model}Model}

Let us consider the system consisting of an inertial Brownian particle moving in an \emph{asymmetric} spatially periodic potential \cite{Hanggi2009}
\begin{equation}
U(x) = \Delta U \left[\sin{\left(2\pi \frac{x}{L} \right)} + 0.25\sin{\left(4\pi \frac{x}{L}\right)}\right]
\end{equation}
and driven out of thermodynamic equilibrium by an external unbiased time-periodic force $A \cos{(\Omega t)}$. The form of dynamics of such a rocking ratchet setup \cite{Cubero2016} taking into account correlations in the surrounding environment and memory effects is given by the following Generalized Langevin Equation \cite{Luczka2005}
\begin{equation} \label{eq:GLE}
    M\ddot{x} + \Gamma\!\!\int_0^t\!\!\! K(t-s)\dot{x}(s)\mathrm{d}s = -U'(x) + A\cos(\Omega t) + \eta(t).
\end{equation}
The dot and prime stand for differentiation with respect to time $t$ and position $x$, respectively. $M$ is the particle mass, $\Gamma$ is the friction coefficient and $K(t)$ is a damping kernel describing via the fluctuation-dissipation theorem \cite{Kubo1966} both the memory as well as correlations of thermal fluctuations $\eta(t)$, namely
\begin{equation}
    \langle \eta(t) \eta(s) \rangle = \Gamma k_B T K(|t-s|).
\end{equation}
Here we choose the simple exponentially decaying kernel characterized by the memory time $\tau_c$
\begin{equation}
    K(t) = \frac{1}{\tau_c}e^{-t/\tau_c},
\end{equation}
for which $\eta(t)$ renders the Orstein-Uhlenbeck process \cite{Luczka2005}.

In order to reduce the number of free parameters and make our analysis independent of a specific setup we introduce the dimensionless position and time variables
\begin{equation}
    \hat{x} = \frac{x}{L}, \quad \hat{t} = \frac{t}{\tau_0},
\end{equation}
where $\tau_0 = \Gamma L^2 / \Delta U$ scales the time needed to travel from the maximum to the minimum of the potential $U(x)$ in the overdamped limit $M\to0$, and $L$ is the spatial period of the ratchet potential. Then we recast Eq.~(\ref{eq:GLE}) to the dimensionless form
\begin{equation} \label{eq:gle}
    m\ddot{\hat{x}} + \int_0^{\hat{t}} \hat{K}(\hat{t}-\hat{s})\dot{\hat{x}}(\hat{s})\mathrm{d}\hat{s} = -\hat{U}'(\hat{x}) + a\cos(\omega \hat{t}) + \hat{\eta}(\hat{t}),
\end{equation}
where
\begin{equation}
    m = \frac{M}{\tau_0 \Gamma},\quad a = \frac{L}{\Delta U}A,\quad \omega = \tau_0\Omega.
\end{equation}
As it is seen in Eq. (\ref{eq:gle}) the dimensionless friction coefficient formally scales to unity, i.e.~$\gamma \equiv 1$. The rescaled potential reads
\begin{equation}
    \hat{U}(\hat{x}) = \sin(2\pi\hat{x}) + \frac{1}{4}\sin(4\pi\hat{x}),
\end{equation}
and the memory kernel is recasted to
\begin{equation} \label{eq:K}
    \hat{K}(\hat{t}) = \frac{1}{\tau}e^{-\hat{t}/\tau},
\end{equation}
where $\tau = \tau_c/\tau_0$ is the dimensionless memory time.
Thermal noise scales as
\begin{equation}
    \hat{\eta}(\hat{t}) = \frac{L}{V_0}\eta(t),
\end{equation}
and its autocorrelation function now reads
\begin{equation}
    \langle \hat{\eta}(\hat{t})\hat{\eta}(\hat{s}) \rangle = D\hat{K}(|\hat{t}-\hat{s}|),
\end{equation}
where $D = k_B T/\Delta U$ is the dimensionless temperature.
From now on we shall analyze only the dimensionless Eq.~(\ref{eq:gle}) and therefore for simplicity we omit the hat over the variables $\hat{x}$, $\hat{t}$, etc.

The Generalized Langevin Equation (\ref{eq:gle}) with an exponentially decaying memory can be transformed into an equivalent set of three differential equations via the Markovian embedding procedure \cite{Luczka2005}
\begin{subequations} \label{eq:gle_set}
\begin{align}
    \dot{x}(t) &= v(t),\\
    m\dot{v}(t) &= z(t) - U'(x) + a\cos(\omega t),\\
    \tau\dot{z}(t) &= -z(t) - v(t) + \xi(t),
\end{align}
\end{subequations}
where $z(t) = \eta(t) - w(t)$ is an auxiliary variable, $w(t)$ represents the integral
\begin{equation}
    w(t) = \int_0^t K(t-s)\dot{x}(s)\mathrm{d}s,
\end{equation}
and $\xi(t)$ is the $\delta$-correlated (white) noise, i.e.
\begin{equation} \label{eq:xi}
    \langle \xi(t) \xi(s) \rangle = 2D\delta(t-s),\quad \langle \xi(t) \rangle = 0.
\end{equation}
The set of equations (\ref{eq:gle_set}) can then be readily solved numerically with standard computational algorithms.

\subsection{Effective mass approach}

On the other hand Eq.~(\ref{eq:gle}) can be approximated with its memoryless counterpart via the effective mass approach. In this scheme the memory effects are reflected purely in the correction to the particle mass \cite{Wisniewski2024-effmass}. The approximate equation reads
\begin{equation} \label{eq:eff_mass}
    m^*\ddot{x}(t) + \dot{x}(t) = -U'(x) + a\cos(\omega t) + \xi(t),
\end{equation}
where 
\begin{equation}
m^* = m - \Delta m(\tau)
\end{equation}
is the effective mass of the particle, and the mass correction $\Delta m(\tau)$ in a general form reads
\begin{equation} \label{eq:dm}
    \Delta m(\tau) = \int_0^\infty tK(t) \mathrm{d}t
\end{equation}
(the dependence of $\Delta m$ on $\tau$ is hidden in the memory kernel $K(t)$). The thermal fluctuations $\xi(t)$ in Eq.~(\ref{eq:eff_mass}) are no longer correlated and are modeled as white noise, see Eq.~(\ref{eq:xi}). This approximation scheme is valid as long as the integral in Eq.~(\ref{eq:dm}) exists and the mass correction is small, that is, $\Delta m(\tau) \ll m$ \cite{Wisniewski2024-effmass}.

For the exponentially decaying memory kernel the mass correction $\Delta m_1$ taking into account only terms up to the first order in the memory time $\tau$ equals
\begin{equation}
    \Delta m_1(\tau) = \tau.
\end{equation}
Moreover, for this specific kernel type it is also possible to derive the correction up to the second order in $\tau$ \cite{Wisniewski2024-2nd_order}, reading 
\begin{equation}	 \label{eq:eff_mass_2nd}
    \Delta m_2(\tau) = \tau\left(1 + \frac{\tau}{m-\tau}\right),
\end{equation}
which improves the accuracy of the effective mass approximation.

\subsection{Quantity of interest}

In this article we analyze how the memory influences direction of the net motion of the particle. For this purpose we define the average velocity of the particle as
\begin{equation}
    \langle v \rangle = \lim\limits_{t\to\infty} \frac{1}{t}\int_0^t \langle \dot{x}(s) \rangle \mathrm{d}s,
\end{equation}
where the triangular brackets $\langle \cdot \rangle$ indicate averaging over the initial conditions and realizations of thermal noise $\eta(t)$. The former is mandatory for the vanishing and low temperature regimes of the system for which its ergodicity may be broken \cite{Spiechowicz2016} and therefore the dynamics might depend on the initial conditions.

\subsection{Methods}

Due to nonlinearity of the potential $U(x)$ and the external driving $a\cos{(\omega t)}$ the set of equations (\ref{eq:gle_set}) cannot be solved analytically in a closed form. For this reason we implemented a weak second-order predictor-corrector algorithm \cite{Platen2010} to simulate the dynamics of the system. To accelerate the calculations we performed the simulations in the CUDA environment on a modern desktop graphics processing unit. This approach allowed us to shorten the computation time by several orders of magnitude \cite{Spiechowicz2015}.

\begin{figure}[t]
    \centering
    \includegraphics[scale=1]{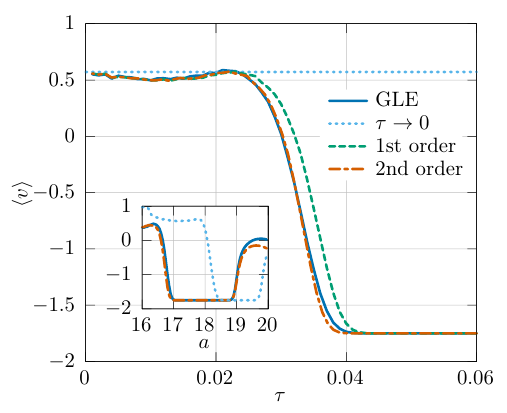}
    \caption{The average velocity of the particle $\langle v \rangle$ as a function of the memory time $\tau$ for the original system (Eq.~\ref{eq:gle}) as well as the effective mass approximation (Eq.~\ref{eq:eff_mass}) in the first and second order. In the inset for $\tau = 0.04$ we show the same characteristic but versus the amplitude $a$ of the external driving force.}
    \label{fig:v_tm}
\end{figure}
\section{\label{sec:results}Results}

The direction of transport in the rocking ratchet system described by Eq.~(\ref{eq:gle}) is very sensitive to changes of the system parameters such as e.g. the driving force amplitude $a$ and frequency $\omega$ \cite{Machura2004}. Therefore here we stick to the following parameter regime
\begin{equation} \label{eq:param}
    \{m = 0.525,\ a = 17,\ \omega = 11,\ D = 10^{-3}\},
\end{equation}
for which in the memoryless limit $\tau \to 0$ the particle average velocity is positive $\langle v \rangle > 0$. 

To illustrate how memory influences the particle transport in this regime, in Fig.~\ref{fig:v_tm} we plot the average velocity $\langle v \rangle$ as a function of the memory time $\tau$ of the kernel $K(t)$. Indeed we can observe that in the limit $\tau \to 0$ the current $\langle v \rangle > 0$. For short memory times $\tau \lesssim 0.03$ the average velocity $\langle v \rangle$ of the particle stays positive, but upon increasing $\tau$ it changes sign, indicating the current reversal effect.
\begin{figure}[b]
    \centering
    \includegraphics[scale=1]{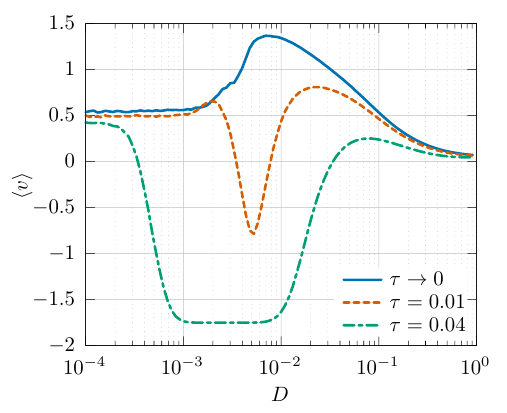}
    \caption{The average velocity $\langle v \rangle$ of the particle versus the temperature $D$ for different memory times $\tau$ and the memoryless limit $\tau\to0$.}
    \label{fig:v_D}
\end{figure}

The origin of this phenomenon can be explained by applying the effective mass approximation. As it is shown in Fig. \ref{fig:v_tm} the current reversal effect emerges also in the memoryless counterpart of the studied system but for the effective $m^* = m - \Delta m(\tau)$ instead of the bare mass $m$. In particular, in the inset we note that the curves corresponding to the original system Eq. (\ref{eq:gle}) with $m = 0.525$ and the effective mass approach with the second order correction Eq. (\ref{eq:eff_mass_2nd}) with $m^* = 0.48$  overlap. It means that the effect of memory is such that the system behaves effectively as it would have a smaller mass for which in the studied parameter regime the current reversal phenomenon arises. 

In Fig.~\ref{fig:v_D} we present the average velocity $\langle v \rangle$ of the particle as a function of the temperature $D$ for different memory times $\tau$. On the one hand, in the deterministic limit $D \to 0$ all curves tend to a positive value. On the other hand, for $D \to \infty$ the fluctuations are so strong that the presence of the potential becomes negligible, and so does its spatial asymmetry, so no net transport occurs regardless of the memory time $\tau$. The difference between the characteristics is visible in the intermediate range of temperatures. In the memoryless limit $\tau \to 0$ the average velocity $\langle v \rangle$ is positive in the whole range of temperatures $D$. When $\tau$ increases, a minimum emerges and the velocity becomes negative for certain values of $D$. This proves that for the system parameters $m$, $a$ and $\omega$ specified in Eq.~(\ref{eq:param}) the current reversal effect can emerge only in the presence of memory, i.e.~for $\tau > 0$.

\begin{figure}[t]
    \centering
    \includegraphics[scale=1]{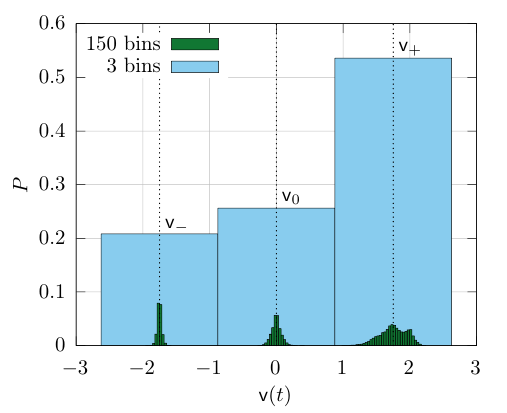}
    \caption{The probability distribution $P(\mathsf{v}(t))$ for the period averaged velocity $\mathsf{v}(t)$ in the memoryless system $\tau \to 0$ for the long time limit calculated from histogram consisting of 150 and 3 bins.}
    \label{fig:histogram}
\end{figure}
The memory-induced current reversal effect can also be explained by analyzing the particle trajectories in the phase space of the system. 
In particular, for each of them
one can calculate the velocity $\mathsf{v}(t)$ averaged over the period of the driving force $\mathsf{T} = 2\pi/\omega$, namely
\begin{equation}
    \mathsf{v}(t) = \frac{1}{\mathsf{T}}\int_{t}^{t+\mathsf{T}} v(s)\mathrm{d}s.
\end{equation}
The average velocity of the particle can then be expressed as the long time limit of the ensemble averaged $\mathsf{v}(t)$ \cite{Jung1993}, i.e.
\begin{equation}
	\langle v \rangle = \lim_{t \to \infty} \langle \mathsf{v}(t) \rangle.
\end{equation}
In Fig.~\ref{fig:histogram} we present the probability distribution $P(\mathsf{v}(t))$ of $\mathsf{v}(t)$ in the memoryless limit $\tau\to0$ calculated for $2^{16} = 65536$ trajectories and time $t=10^4$ for which the asymptotic state of the system has already been reached. Velocity $\mathsf{v}(t)$ is distributed around three values corresponding to average velocities of the particle for three attractors present in the deterministic limit of vanishing thermal noise $D \to 0$ (not depicted). Two $\mathsf{v}_{\pm} = \pm \omega/(2\pi)$ of these values render running solutions in which the particle covers one spatial period during one period of the driving force. The third one $\mathsf{v}_0 = 0$ indicates the locked state in which the particle motion is limited to one potential well and no net transport is observed.

\begin{figure}[t]
    \centering
    \includegraphics[scale=1]{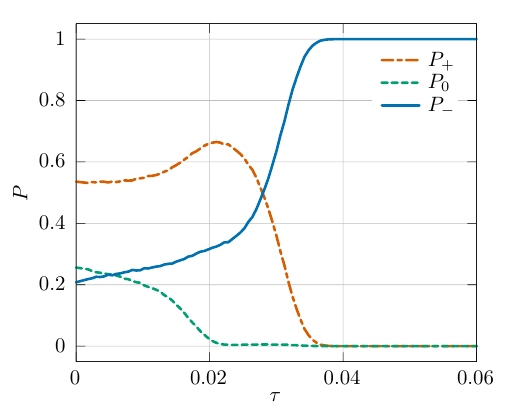}
    \caption{The probabilities $P_\pm$ and $P_0$ for the emergence of the states $\mathsf{v}_\pm = \pm \omega/(2\pi)$ and $\mathsf{v}_0 = 0$, respectively, versus the memory time $\tau$.}
    \label{fig:probability}
\end{figure}
For thermal noise assisted dynamics the particle can randomly switch between these solutions, however, in the long time limit the probability $P(\mathsf{v}(t))$ must be a time-invariant measure. In Fig. \ref{fig:probability} we present the probabilities $P_\pm$ and $P_0$ corresponding to the three observed states $\mathsf{v}_\pm = \pm \omega/(2\pi)$ and $\mathsf{v}_0$, respectively, as a function of the memory time $\tau$. The quantities $P_\pm$ and $P_0$ can be easily determined directly from the analogues of Fig.~\ref{fig:histogram} for different $\tau$.
In the memoryless limit $\tau \to 0$ most of the system trajectories render the positive velocity $\mathsf{v}_+$ and therefore in this regime the particle current $\langle v \rangle$ is positive as well. Upon increasing the memory time $\tau$ the state $\mathsf{v}_-$ quickly becomes more populated and later it saturates at unity $P_- = 1$. It means that the probability of escaping the state $\mathsf{v}_-$ becomes negligible. This explains why the particle current reversal phenomenon emerges. It is rooted in the memory-induced dynamical localization effect \cite{Spiechowicz2017b, Spiechowicz2019} in which all system trajectories possess the negative velocity $\mathsf{v}_-$ when the correlation time $\tau$ is increased.

\section{\label{sec:conclusions}Conclusions}

In this paper we revisited the everlasting problem of the role of memory in dynamical systems. In particular we considered a model of a Brownian ratchet, namely a Brownian particle in an asymmetric spatially periodic potential immersed in a correlated thermal bath and driven by an external force. Our primary interest was the net transport in such a system characterized by an average velocity of the Brownian particle in the long time regime. 

We showed that the direction of the transport can be reversed upon increasing the memory time indicating the memory-induced current reversal effect. Moreover, we showed that in the memoryless limit the average velocity of the particle stays positive regardless of the temperature of the system. This confirms that in the studied parameter regime the current reversal effect can be observed only in the presence of memory.

We explained this phenomenon in two ways. Firstly, we applied the effective mass approach to our system, in which the presence of short memory is represented as a correction to the particle mass. It turned out that the impact of memory is such that the system behaves effectively as it would have a mass smaller than the actual one. In the studied parameter regime the current reversal phenomenon arises for this reduced effective mass while it ceases to exist for the physical one corresponding to the memoryless system.

Secondly, we calculated the probability for the particle to reside in each of three states for the period-averaged velocity present in the deterministic counterpart of the system. Upon increasing the memory time, occupation of the states changes so that finally all trajectories possess the negative velocity. Such memory-induced dynamical localization of the trajectories in the velocity subspace is the mechanism underlying the observed current reversal effect.

Our results reveal new aspects of the influence of memory on the dynamics of Brownian ratchet, and can be corroborated experimentally in variety of setups embodying this system.

\section*{Acknowledgments}
This work has been supported by the National Science Centre (NCN) under Grant No. 2022/45/B/ST3/02619 (J.S.).

\bibliographystyle{apsrev}
\bibliography{bibliography.bib}




\end{document}